\documentclass[copyright,creativecommons]{eptcs}
\providecommand{\event}{E. Formenti (ed.): AUTOMATA \& JAC 2012} 
\usepackage[strings]{underscore}  

\newcommand*{\set}[1]{\{#1\}}

\newcommand*{\plusc}{+_{\ldots}}

\newcounter{tempcount}
\newcommand{\Hn}[1]{{e}^{\Theta(\sqrt{{#1}\cdot\ln{#1}})}}

\newcommand*{\logspace}{\ensuremath{\mathsf{L}}}
\newcommand*{\nlogspace}{\ensuremath{\mathsf{NL}}}

\newcommand*{\gap}{\ensuremath{\mathsf{GAP}}}

\newcommand*{\dgap}{\ensuremath{D_{\gap}}}

\newcommand*{\lpoly}{\ensuremath{\mathsf{L/poly}}}

\newcommand{\unarygap}{\ensuremath{\mathsf{UGAP}}}

\newcommand{\Unarygap}[1]{\ensuremath{\unarygap\!_{#1}}}

\newcommand{\uencoding}[1]{\ensuremath{\langle{#1}\rangle_{\!_1}}}

\newcommand*{\tw}{\textsc{2}}
\newcommand*{\twdfa}{\tw\textsc{dfa}}
\newcommand*{\twdfas}{\tw\textsc{dfa}s}

\newcommand*{\twnfa}{\tw\textsc{nfa}}
\newcommand*{\twnfas}{\tw\textsc{nfa}s}

\newcommand*{\twonfas}{\tw\textsc{ofa}s}

\newcommand*{\ow}{\textsc{1}}
\newcommand*{\owdfa}{\ow\textsc{dfa}}
\newcommand*{\owdfas}{\ow\textsc{dfa}s}

\newcommand*{\ownfa}{\ow\textsc{nfa}}
\newcommand*{\ownfas}{\ow\textsc{nfa}s}

\newtheorem{theorem}{Theorem}[section]
\newtheorem{lemma}{Lemma}[section]

\title{Two-Way Finite Automata: Old and Recent Results}
\author{Giovanni Pighizzini
\institute{Dipartimento di Informatica}
\institute{Universit\`a degli Studi di Milano -- Italia}
\email{pighizzini@di.unimi.it}
}
\def\titlerunning{Two-Way Finite Automata: Old and Recent Results}
\def\authorrunning{G.\ Pighizzini}
\begin{document}
\maketitle

\begin{abstract}
The notion of  \emph{two-way automata} was introduced at the very beginning of automata theory.
In 1959, Rabin and Scott~\cite{RS59} and, independently, 
Shepherdson~\cite{Sh59}, proved that these models, 
both in the deterministic and in the nondeterministic versions, have the same power of 
one-way automata, namely, they characterize the class of \emph{regular languages}.
In 1978, Sakoda and Sipser~\cite{SS78} posed the question of the cost, in the number of the states, 
of the simulation of one-way and two-way nondeterministic automata by  two-way deterministic automata. 
They conjectured that these costs are exponential. In spite of all attempts to solve it, 
this question is still open.
In the last ten years the problem of Sakoda and Sipser was widely reconsidered and many new results 
related to it have been obtained. In this work we discuss some of them. 
In particular, we focus on the restriction to the unary case and on the connections with open 
questions in space complexity.
\end{abstract}

\section{Introduction and Preliminaries}

Finite state automata are usually presented as devices which are able to recognize input
strings using a fixed amount of memory, implemented using a finite state control (see, e.g.,~\cite{HU79}).
The input string is written on a read-only tape, which is scanned by an input head.
In the basic model the input head is moved only {}from left to right.
For this reason the model is also called \emph{one-way finite automaton}. 
It can be defined in the
deterministic and the nondeterministic versions (\owdfa\ and \ownfa, respectively). It is
well known that both of them share the same recognition power, i.e., 
they characterize the class of regular languages.
However, nondeterministic finite automata can be exponentially smaller.
In fact, each $n$-state \ownfa\ can be simulated by an equivalent \owdfa\ with $2^n$ states
and this cost cannot be reduced~\cite{Lup66,MF71,Moo71}.

\emph{What happens if we allow to move the input head in both directions?}

In spite of this additional feature, the resulting models, which are called
\emph{two-way finite automata}, have the same computational power as 
one-way automata, i.e., they still characterize the class of regular languages,
as independently proved by Rabin and Scott~\cite{RS59} and by Shepherdson~\cite{Sh59}, at the
beginning of automata theory.
However, {}from the point of view of the size (measured in terms of states) the situation is
different. We still do not have a complete picture of the relationships between the
sizes of different variants of finite automata.

By an analysis of the constructions given in~\cite{RS59,Sh59}, 
it turns out that the simulations of $n$-state two-way nondeterministic finite automata (\twnfas, for short)
and  $n$-state two-way deterministic finite automata (\twdfas, for short) by
\owdfas\ can be done with a number of states exponential in a polynomial in $n$.
Furthermore, a lower bound exponential in $n$ follows {}from the simulation of
\ownfas\ by \owdfas.
The exact bound for the simulation of \twnfas\ by \ownfas\ has been found in~\cite{Kap05}.

\emph{The costs of the simulations of \ownfas\ by \twdfas\ and of \twnfas\ by \twdfas\  are still unknown.} 
The problem of stating them was raised in 1978 by Sakoda and Sipser~\cite{SS78},
with the conjecture that they are not polynomial.
In spite of all attempts to solve it, this problem is still open.

In the last decade several new results related to the Sakoda and Sipser question have been
discovered. In this paper we discuss some of them (mainly with respect to
the question of \twnfas\ versus \twdfas) besides some older results
in this area.

\subsection*{Technical Issues}

We will keep the presentation at an informal level, trying to avoid, as much as
possible, technical details. For this reason we do not give a formal definition
of the main model we are interested in, but we  just present an informal description.

We assume that the reader is familiar with standard notions concerning
finite state automata, as presented for instance in~\cite{HU79}.
We denote by~$\Sigma$ the \emph{input alphabet},
by~$\Sigma^*$ the set of all strings over $\Sigma$, and by $\Sigma^n$ the set of
strings of length $n$, where $n\geq 0$ is an integer.
The length of a string $w\in\Sigma^*$ will be denoted by $|w|$.

A computation of a one-way automaton starts on the leftmost input symbol in the initial state;
at each step the input head is moved one position to the right; the computation ends
immediately after the execution of the move which reads the rightmost input symbol.
For two-way automata slightly different definitions are given in the literature.
We skip technical details and we emphasize the main features.
\begin{itemize}
\item First of all, we assume that the input string is surrounded on the input tape by two special
  symbols, $\vdash,\dashv\notin\Sigma$, called, respectively, the \emph{left} and the 
  \emph{right endmarker}. Hence, if the input is $w\in\Sigma^*$, then the input tape contains
  $\vdash w\dashv$.
\item To present recognition algorithms, sometimes we need to number input cells.
  So, we assume that on input $w$ the cells are numbered {}from $0$ to $|w|+1$, where
  cells $0$ and $|w|+1$ contain the endmarkers, and the remaining cells contain ``real''
  input symbols. The input head cannot violate the endmarkers.
\item The computation starts in a designed initial state with the head scanning the first ``real'' 
  input symbol, i.e., on cell $1$. Sometimes it is more convenient to start {}from cell $0$. It should
  be clear that this does not significantly change the model.
\item To reflect the acceptance condition for one-way automata, we can stipulate that
  a string is accepted by a two-way automaton if and only if there is a computation which reaches the right 
  endmarker in a final state. However, this condition can be slightly modified by considering acceptance
  on the left endmarker or just on one endmarker.
  
  A different possibility is to state that a string is accepted if and only if there is a computation
  which reaches a final state, regardless the input head position.
  
  Further variants are possible.  
  It should be clear that all these variants are equivalent. Adding one or two states, we can
  easily convert a two-way automaton with an acceptance condition into another one with a different
  acceptance condition.
  For this reason, here we do not fix any particular acceptance condition.
  
\item The transition function can be defined by allowing only moves to the left and to the
  right or even allowing stationary moves, i.e., transitions that keep the head on the same input cell.
  Even this possibility does not significantly change the model and the number of  states.

\item We point out that a two-way automaton can enter into a loop. In this case the computation is
  rejecting.

\item When we say that a two-way automaton $A$ has $f(n)\plusc$ states, we mean that 
  $A$ has $f(n)+c$ states, where $c$ is a small constant
  (in all examples $c<10$ is enough). This constant can slightly change depending on the
  choice of the initial configuration, of the acceptance condition, and of the possibility of
  stationary moves.

\item An \emph{head reversal} is any change of the input head direction, i.e.,
  a two-way automaton makes one head reversal when after a sequence of transitions moving the
  head to the right it make a transition moving the head to the left or vice versa.
  Stationary moves are not taken into account to compute head reversals. For instance
  a sequence of two moves to the right, one stationary move, one move to the right,
  one stationary move again and one move to the left contains just one head reversal.

\end{itemize}

\section{Two Examples}

Let us start by considering the following family of languages
\[
I_n=(a+b)^*a(a+b)^{n-1}~,
\]
namely, for each integer $n>0$, $I_n$ is the set of strings
whose $n$th symbol {}from the right is an $a$.
This is a classical example used to present the
optimality of the subset construction (actually, this very simple example
does not achieve exactly the optimality, but it is very close to it).
In particular, for each $n\geq 1$, we can prove the following:
\begin{itemize}
\item The language $I_n$ is accepted by the  \ownfa\ with $n+1$ states in Figure~\ref{f:In}.
\begin{figure}[hbt]
\begin{center}
\setlength{\unitlength}{0.5cm}
\begin{picture}(24,4)(7,3)
  \put(9,5){\circle{2}}
  \put(9,5){\makebox(0,0){$q_0$}}
  \put(14,5){\circle{2}}
  \put(14,5){\makebox(0,0){$q_1$}}
  \put(19,5){\circle{2}}
  \put(19,5){\makebox(0,0){$q_2$}}
  \put(24,5){\circle{2}}
  \put(24,5){\makebox(0,0){$q_3$}}
  \put(30,5){\circle{2}}
  \put(30,5){\makebox(0,0){$q_n$}}
  \put(30,5){\circle{1.6}}

	\put(7.1,6.9){\vector(1,-1){1.1}}
  
  \put(10,5){\vector(1,0){3}} 
  \put(11.5,5.5){\makebox(0,0){$a$}}
  \put(15,5){\vector(1,0){3}} 
  \put(16.5,5.5){\makebox(0,0){$a,b$}}
  \put(20,5){\vector(1,0){3}} 
  \put(21.5,5.5){\makebox(0,0){$a,b$}}
  \multiput(25,5)(0.2,0){15}{\line(1,0){0.1}}
  \put(28,5){\vector(1,0){1}} 
  \put(27.5,5.5){\makebox(0,0){$a,b$}}

  \put(7.5,4){\oval(3,2)[br]}
  \put(7.5,4){\oval(2,2)[l]}
  \put(7.5,5){\vector(1,0){0.5}}
  \put(9.7,3.5){\makebox(0,0){$a,b$}}
  
\end{picture}
\caption{A \ownfa\ accepting the language $I_n=(a+b)^*a(a+b)^{n-1}$.}\label{f:In}
\end{center}
\end{figure}
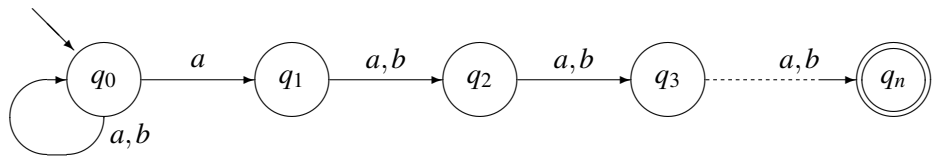
\item Each \owdfa\ accepting $I_n$ requires $2^n$ states.
  Intuitively, this can be proved by observing that in order to accept
  the language $I_n$, a \owdfa\ needs to remember the last $n$ input symbols.
  It is a standard exercise to depict a \owdfa\ matching this lower bound.
\item The language $I_n$ is accepted by a \twdfa\ with $n\plusc$ states which reverses
  its input head just one time during each computation. The automaton,
  firstly scans the input {}from left to right, only to reach the
  right endmarker. Then it moves $n$ positions to the left,
  finally checking whether or not the reached input cell contains the
  symbol $a$.
\end{itemize}
This simple example emphasizes that the possibility 
of moving the input head in both directions can drammatically reduce
the size of deterministic automata. In particular, in this case
one reversal is enough to reduce an automaton
of exponential size in $n$ to an automaton of linear size.

We can also observe that the language $I_n$ is accepted by
a \ownfa\ and a \twdfa\ having approximatively the same size.
So the example could suggest the possibility of replacing
the nondeterminism in one-way automata by two-way motion.

We now present a more elaborated variant of this example
which will be also useful to discuss some restricted versions of
two-way automata considered in the literature.
For each $n>0$, let us consider the language
\[
L_n=(a+b)^*a(a+b)^{n-1}a(a+b)^*~.
\]
In this case we ask that each string in the language contains
two letters $a$'s with $n-1$ symbols in between.

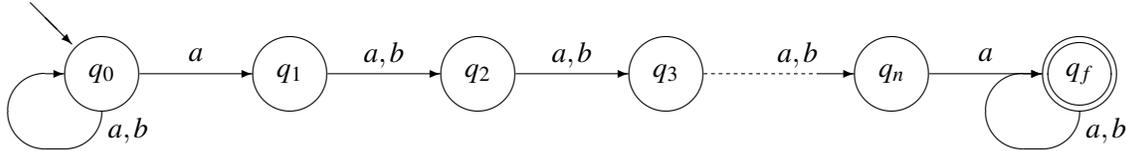
\begin{figure}[hbt]
\begin{center}
\setlength{\unitlength}{0.5cm}
\begin{picture}(29,4)(7,3)
  \put(9,5){\circle{2}}
  \put(9,5){\makebox(0,0){$q_0$}}
  \put(14,5){\circle{2}}
  \put(14,5){\makebox(0,0){$q_1$}}
  \put(19,5){\circle{2}}
  \put(19,5){\makebox(0,0){$q_2$}}
  \put(24,5){\circle{2}}
  \put(24,5){\makebox(0,0){$q_3$}}
  \put(30,5){\circle{2}}
  \put(30,5){\makebox(0,0){$q_n$}}
  \put(35,5){\circle{2}}
  \put(35,5){\makebox(0,0){$q_f$}}
  \put(35,5){\circle{1.6}}

  \put(7.1,6.9){\vector(1,-1){1.1}}

  \put(10,5){\vector(1,0){3}} 
  \put(11.5,5.5){\makebox(0,0){$a$}}
  \put(15,5){\vector(1,0){3}} 
  \put(16.5,5.5){\makebox(0,0){$a,b$}}
  \put(20,5){\vector(1,0){3}} 
  \put(21.5,5.5){\makebox(0,0){$a,b$}}
  \multiput(25,5)(0.2,0){15}{\line(1,0){0.1}}
  \put(28,5){\vector(1,0){1}} 
  \put(27.5,5.5){\makebox(0,0){$a,b$}}
  \put(31,5){\vector(1,0){3}} 
  \put(32.5,5.5){\makebox(0,0){$a$}}

  \put(7.5,4){\oval(3,2)[br]}
  \put(7.5,4){\oval(2,2)[l]}
  \put(7.5,5){\vector(1,0){0.5}}
  \put(9.7,3.5){\makebox(0,0){$a,b$}}

  \put(33.5,4){\oval(3,2)[br]}
  \put(33.5,4){\oval(2,2)[l]}
  \put(33.5,5){\vector(1,0){0.5}}
  \put(35.7,3.5){\makebox(0,0){$a,b$}}
  
\end{picture}
\caption{A \ownfa\ accepting the language $L_n=(a+b)^*a(a+b)^{n-1}a(a+b)^*$.}\label{f:Ln}
\end{center}
\end{figure}

The language $L_n$ can be easily accepted by the \ownfa\ with $n+2$ states
in Figure~\ref{f:Ln}.

What about acceptance of $L_n$ by one-way and two-way
\emph{deterministic} automata?

Let us start by studying acceptance in the one-way case.
The idea is very similar to the one outlined for the language $I_n$.

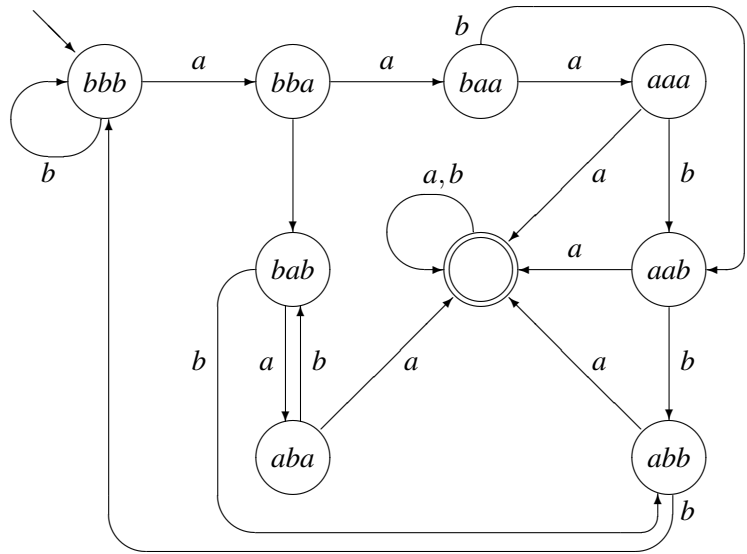
\begin{figure}[hbt]
\begin{center}
\setlength{\unitlength}{0.5cm}
\begin{picture}(19,15)(0,0)
  \put(0.1,14.9){\vector(1,-1){1.1}}

  \put(2,13){\circle{2}}
  \put(2,13){\makebox(0,0){$bbb$}}
  \put(7,13){\circle{2}}
  \put(7,13){\makebox(0,0){$bba$}}
  \put(12,13){\circle{2}}
  \put(12,13){\makebox(0,0){$baa$}}
  \put(17,13){\circle{2}}
  \put(17,13){\makebox(0,0){$aaa$}}
  \put(3,13){\vector(1,0){3}} 
  \put(4.5,13.5){\makebox(0,0){$a$}}
  \put(8,13){\vector(1,0){3}} 
  \put(9.5,13.5){\makebox(0,0){$a$}}
  \put(13,13){\vector(1,0){3}} 
  \put(14.5,13.5){\makebox(0,0){$a$}}
  \put(7,12){\vector(0,-1){3}}
  \put(13.4,14){\oval(2.8,2)[tl]}
  \put(13.4,14){\oval(11.2,2)[tr]}
  \put(18.5,9){\oval(1,2)[br]}
  \put(19,9){\line(0,1){5}}
  \put(18.55,8){\vector(-1,0){0.55}}
  \put(11.5,14.5){\makebox(0,0){$b$}}
  \put(17,12){\vector(0,-1){3}}
  \put(17.5,10.6){\makebox(0,0){$b$}}
  \put(16.25,12.25){\vector(-1,-1){3.5}}
  \put(15.15,10.5){\makebox(0,0){$a$}}

  \put(0.5,12){\oval(2.8,2)[br]}
  \put(0.5,12){\oval(2,2)[l]}
  \put(0.5,13){\vector(1,0){0.5}}
  \put(0.5,10.6){\makebox(0,0){$b$}}

  \put(7,8){\circle{2}}
  \put(7,8){\makebox(0,0){$bab$}}
  \put(12,8){\circle{2}}
  \put(12,8){\circle{1.6}}
  \put(17,8){\circle{2}}
  \put(17,8){\makebox(0,0){$aab$}}
  \put(6.8,7.05){\vector(0,-1){3.1}}
  \put(6.3,5.5){\makebox(0,0){$a$}}
  \put(17,7){\vector(0,-1){3}}
  \put(17.5,5.6){\makebox(0,0){$b$}}
  \put(16,8){\vector(-1,0){3}}
  \put(14.5,8.5){\makebox(0,0){$a$}}
  \put(7.75,3.75){\vector(1,1){3.5}}
  \put(10.15,5.5){\makebox(0,0){$a$}}
  \put(16.25,3.75){\vector(-1,1){3.5}}
  \put(15.15,5.5){\makebox(0,0){$a$}}
  \put(10.4,9){\oval(2.8,2)[tr]}
  \put(10.5,9){\oval(2,2)[l]}
  \put(10.5,8){\vector(1,0){0.5}}
  \put(11,10.5){\makebox(0,0){$a,b$}}
  
  \put(6,4.5){\oval(2,7)[l]}
  \put(6,1){\line(1,0){9.7}}
  \put(16.7,1.5){\vector(0,1){0.55}}
  \put(15.7,1.5){\oval(2,1)[br]}
  \put(4.5,5.6){\makebox(0,0){$b$}}

  \put(7,3){\circle{2}}
  \put(7,3){\makebox(0,0){$aba$}}
  \put(17,3){\circle{2}}
  \put(17,3){\makebox(0,0){$abb$}}
  \put(7.2,3.95){\vector(0,1){3.1}}
  \put(7.7,5.6){\makebox(0,0){$b$}}
  \put(17.5,1.6){\makebox(0,0){$b$}}
  \put(16.1,2){\oval(2,3)[br]}
  \put(3.1,0.5){\line(1,0){13}}
  \put(3.1,2){\oval(2,3)[bl]}
  \put(2.1,2){\vector(0,1){10}}

\end{picture}
\caption{The \owdfa\ $A_3$ accepting the language $L_3=(a+b)^*a(a+b)^{2}a(a+b)^*$.}\label{f:L3}
\end{center}
\end{figure}

We can build an automaton $A_n$ which remembers in its final control
the last $n$ input symbols. Hence, when in the state corresponding
to $\sigma_1\sigma_2\ldots\sigma_n$ a new input symbol $\gamma$ is read, 
the automaton moves to the state corresponding to $\sigma_2\ldots\sigma_n\gamma$.
However, in the case $\sigma_1=\gamma=a$ the automaton moves
to its only final state, where it loops on each input symbol.
In Figure~\ref{f:L3}, the automaton $A_3$ accepting the
language $L_3$ is represented. Notice that with this strategy
the resulting \owdfa\ has $2^n+1$ states.

\noindent
We can show that each automaton $A_n$ is minimal. This can be done
by using classical distinguishability arguments (see, e.g.,~\cite{HU79}) along the following
lines:
\begin{itemize}
\item Each two pairwise different strings $x,y$ of length $n$
  are distinguishable. To prove this it is enough to consider
  the string $b^{i-1}a$, where $i$, $1\leq i\leq n$,
  is the index of the leftmost letter different in $x$ and~$y$, 
  and to verify that exactly one string between
  $xb^{i-1}a$ and $yb^{i-1}a$ belongs to $L_n$.
\item Each string of length $n$ does not belong to $L_n$
  and, hence, it is distinguishable {}from $a^{n+1}$ which
  belongs to $L_n$.
\item Hence, the $2^{n}+1$ strings in the set $\Sigma^n\cup\set{a^{n+1}}$ are pairwise 
  distinguishable for $L_n$. As a consequence, $2^{n}+1$ is a lower bound for the number 
  of states of each \owdfa\ accepting $L_n$. This lower bound matches the number of the states
  of the automaton $A_n$ above described.
\end{itemize}

\noindent
Now, we discuss a different strategy to accept $L_n$ using a two-way automaton.
In the following let $w=w_1w_2\cdots w_m$, with $w_i\in\set{a,b}$, $i=1,\ldots,m$, $m\geq 0$,
be the input string for which we want to check the membership to $L_n$.

\begin{enumerate}
\item[(i)]\emph{Na\"{\i}f algorithm}\\
  To decide whether a string $w\in\Sigma^*$ belongs to $L$, for 
  $i=1,\ldots,|w|-n$ we check if both symbols in positions $i$ and $i+n$ are $a$'s.
  The input is accepted if for at least one $i$ the condition is satisfied.
  This algorithm can be implemented by a \twdfa\ that to move {}from position
  $i$ to position $i+n$ counts $n$ positions forward, and then counts $n-1$ positions 
  backward to reach position $i+1$. Furthermore, when moving {}from position $i$ to
  position $i+n$, the automaton needs to remember whether or not the symbol in position $i$
  is $a$.
  This leads to a \twdfa\ with $O(n)$ states which moves the input head 
  along a zig-zag trajectory. 
    
\item[(ii)]\emph{An improved algorithm}\\
  It is immediate to observe that the na\"{\i}f algorithm can be improved.
  First, when the symbol $w_i$ is $b$, we do not need to inspect the symbol $w_{i+n}$.
  Second, when a position $i$ is found such that both symbols $w_i$ and $w_{i+n}$ are $a$'s,
  the automaton can
  accept without checking the remaining positions. This leads to an algorithm which uses
  no more than $2n\plusc$ states. 

\item[(iii)]\emph{A different strategy: head reversals only at the endmarkers}\\
  We can describe a different algorithm to recognize $L_n$, which is implemented 
  by a \twdfa\ performing head reversal \emph{only} when the input head is
  visiting the endmarkers. Hence, in this algorithm a computation is a sequence of 
  left-to-right and right-to-left traversals of the input string, which
  are also called \emph{sweeps}.
  
  We give an informal description of the algorithm:
  \begin{itemize}
  \item The automaton performs at most $n$ sweeps {}from left to right, interleaved
  with sweeps {}from right to left.
  \item In the $i$th sweep {}from left to right, $1\leq i\leq n$, the automaton starting
  {}from the cell $i$, inspects the contents of cells
  $i,~i+n,~i+2\cdot n,~i+3\cdot n,~\ldots~$, in order to check if two
  of them which are consecutive in this list (i.e., cells $i+j\cdot n$
  and $i+(j+1)\cdot n$, for some $j\geq 0$) contain the symbol $a$.
  If this happens then the automaton stops and accepts.
  
  To locate the cells that must be inspected, a counter $c_{\rightarrow}$ modulo $n$ is
  kept in the finite control. This counter can be implemented using $n$ states.
  However, the automaton needs to remember the content of the last inspected cell.
  This doubles the number of the states.
  
	\item When in the $i$th scan {}from left to right, the right endmarker is reached, 
	there are two possibilities.
	If $i<n$ then the automaton makes a sweep {}from right to left, in order to prepare the 
	$(i+1)$th scan {}from left to right. If $i=n$ then the automaton stops and rejects.
\end{itemize}
	
This strategy can be implemented with $O(n^2)$ states, by keeping track in
	the finite control of the counter $i$, and by using $2n$ states for each sweep
	{}from left to right, and just one state for each sweep {}from right to left.

We can reduce the number of states to $O(n)$ by avoiding to store the counter $i$ for sweeps. 
To this aim, also during sweeps {}from right to left we count the input length modulo $n$, by
introducing another counter $c_{\leftarrow}$. After the $i$th sweep {}from left to right, the sweep {}from right to 
left starts by assigning to the counter $c_{\leftarrow}$ a value which depends on the current value of $c_{\rightarrow}$.
In this way, at the end of the traversal {}from right to left, when the left endmarker is
reached again, {}from the value of~$c_{\leftarrow}$ it is possible to reconstruct the value of $i$, in order
to prepare the next sweep.
\end{enumerate}

\section{Restricted Models}

We now briefly present and discuss some restricted variants of two-way automata that have
been considered in the literature.

\subsubsection*{Oblivious Automata}
In the na\"if algorithm (i) we described to recognize language $L_n$, we can
observe that for all the inputs of the same length $m$ the ``trajectory'' of the head 
during the computation is the same, i.e., the position of the
input head at the time $t$ does not depend on the input content, but only on its length.
A \twdfa\ with this property is called \emph{oblivious}.

\subsubsection*{Sweeping Automata}
A two-way automaton  performing head reversal \emph{only}  when the input head is
visiting the endmarkers is called \emph{sweeping automaton}. This notion
has been studied by Sipser~\cite{Sip80}.
In particular, for the language $L_n$ above described, the recognition strategy
(iii) is based on a sweeping \twdfa. 
  
\subsubsection*{Rotating Automata}
The method (iii) suggests another model, called \emph{rotating automata}~\cite{KKM12}, which
now we briefly mention.
A computation of a rotating automaton is a sequence of left-to-right scans of the
input. In particular, when the right end of the input is reached, the computation
continues on the leftmost input symbol. In other words, we can imagine the input tape 
as circular, with a special cell containing a marker and connecting the end with the
beginning of the tape. With a trivial transformation which doubles the number of the states,
each rotating automaton can be transformed into an equivalent sweeping automaton.

The reader can verify that languages $I_n$ and $L_n$ can be accepted by
rotating automata with $O(n)$ states.

\subsubsection*{Outer Nondeterministic Automata}
All the above mentioned models are defined by restricting the movement of the
input head. A different kind of restriction has been recently considered in~\cite{GGP12,KP12a},
by introducing \emph{outer nondeterministic automata} (\twonfas).
In these models nondeterministic choices can be taken \emph{only} when the
input head is scanning the endmarkers. Hence, the transition on ``real'' input symbols
are deterministic. This model does not have any restriction on head reversals, i.e.,
\twonfas\ can change the direction of the input head at each position.

The deterministic algorithm (iii) for accepting $L_n$ can be easily transformed
in an algorithm for a (degenerate) outer nondeterministic automaton.
At the first step the automaton guesses an integer $i$, with $1\leq i\leq n$, and then
it simulates the $i$th sweep {}from left to right described in algorithm (iii),
rejecting if the right endmarker is reached without finding two cells
$i+j\cdot n$ and $i+(j+1)\cdot n$, both containing the symbol $a$. This can be implemented
just choosing the initial value of the counter $c_{\rightarrow}$ in a nondeterministic way, 
at the beginning of the computation with the head
on the left endmarker.

\subsubsection*{Few Reversal Automata}
All the models above discussed  are defined by introducing structural restrictions.
In the next model the restriction is of a different kind. On each computation we count
the number of reversals of the input head during the computation.
A \twdfa\ is said to be \emph{few reversals} if the number of head reversals is sublinear
with respect to the input length, i.e., it is $o(m)$, where $m$ is the length of the
input. It has been recently proved that a \twdfa\ with $o(m)$ reversals is actually
a \twdfa\ with $O(1)$ reversals, i.e., each few reversal \twdfa\ can make only a
number of reversals which is ultimately bounded by a constant~\cite{KP12b}.

Notice that the algorithm (i) above described clearly uses a number of reversals
which is linear in the length of the input.
Even the algorithm (ii) uses a linear number of reversals (consider, e.g.,
inputs of the form $a^nb^na^nb^n\ldots a^nb^n$).
On the other hand, in the algorithm (iii) the number of reversals is bounded
by $2n-1$, which is a constant with respect to the input length.

In the nondeterministic case we can have several computations for a same input
string. For this reason we can measure head reversals in different ways. For example,
we can consider reversals in \emph{all computations}, or only in \emph{all accepting
computations}, or just in \emph{one accepting computation}. This can lead to different
notions of few reversal \twnfas\ (something similar is well known in space complexity,
where different space notions have been considered, see, e.g.,~\cite{Me08}).

\subsubsection*{Unambiguous Automata}
This is a well known classical notion: a nondeterministic automaton is \emph{unambiguous}
if and only if for each input string there is at most one accepting computation.
While the \ownfa\ above described to recognize $I_n$ is unambiguous, it can be easily
seen that the \ownfa\ $A_n$ accepting $L_n$ can have many accepting computation for a same
input string, i.e., it is ambiguous.

\section{Restrictions on the Simulating Machines}

As already mentioned in the introduction, the Sakoda and Sipser question asks
the costs, in states, of the simulations of \ownfas\ and \twnfas\ by \twdfas.
Separations have been obtained by considering restrictions on the target machines.
In particular, the simulations of $n$-state \ownfas\ (and hence also \twnfas)
by sweeping, oblivious, and few reversal automata  require exponentially many 
states.\footnotemark\setcounter{tempcount}{\thefootnote}

Note that all above restrictions are related to the movement of the input head.

\begin{figure}[t]
\begin{center}
\setlength{\unitlength}{0.5cm}
\begin{picture}(36,15)(0,1)

	\put(18,15){\makebox(0,0)[b]{\ownfa}}
	\put(11,8.5){\makebox(0,0)[r]{\small oblivious}}
	\put(18,8.5){\makebox(0,0){\small sweeping}}
	\put(25,8.5){\makebox(0,0)[l]{\small few reversals}}
	\put(18,2){\makebox(0,0)[t]{\twdfa}}
	
	\put(16,2.5){\vector(-1,1){5}} 
	\put(13.5,4.5){\makebox(0,0)[r]{\small\cite{KMP12}}}
	\put(18,2.5){\vector(0,1){5}}  
	\put(17.9,4.5){\makebox(0,0)[r]{\small\cite{Ber80,Mic81,Sip80}}}
	\put(20,2.5){\vector(1,1){5}}  
	\put(22.4,4.5){\makebox(0,0)[l]{\small\cite{Kap11b}}}

	\put(11.2,9.5){\vector(1,1){5}}  
	\put(13.4,12.5){\makebox(0,0)[r]{\small\cite{HS03}\footnotemark[\thetempcount]}}
	\put(15.7,14.5){\vector(-1,-1){5}}  
	\put(14.5,12.5){\makebox(0,0)[l]{\small\cite{KMP12}}}
	\put(18.2,14.5){\vector(0,-1){5}}   
	\put(18.3,11.5){\makebox(0,0)[l]{\small\cite{Sip80}}}
	\put(17.8,9.5){\vector(0,1){5}}   
	\put(17.7,11.5){\makebox(0,0)[r]{\small\cite{KMP12}}}
	\put(17.7,12.5){\makebox(0,0)[r]{\small(d)}}
	\put(20.3,14.5){\vector(1,-1){5}}   
	\put(22.6,12.5){\makebox(0,0)[l]{\small\cite{Kap11b}}}
	\put(24.8,9.5){\vector(-1,1){5}}   
	\put(21.5,12.5){\makebox(0,0)[r]{\small(c)}}
	
	\put(11.5,8.7){\vector(1,0){4.5}}  
	\put(13.75,8.8){\makebox(0,0)[b]{\small\cite{KMP12}}}

	\put(24.5,8.7){\vector(-1,0){4.5}} 
	\put(22.25,8.8){\makebox(0,0)[b]{\small\cite{Kap11b}}}

  \multiput(20,8.3)(0.4,0){10}{\line(1,0){0.2}}
	\put(24,8.3){\vector(1,0){0.5}}  
	\put(22.25,8.2){\makebox(0,0)[t]{\small(b)}}

  \multiput(12.2,8.3)(0.4,0){10}{\line(1,0){0.2}}
	\put(12,8.3){\vector(-1,0){0.5}}  
  \put(13.75,8.2){\makebox(0,0)[t]{\small(a)}}

\end{picture}
\caption{An arrow {}from a class $A$ of machines to a class $B$ denotes an exponential
separation, i.e., the state cost of the simulation of machines in the class
$A$ by machines in the class $B$ can be exponential.
A dashed arrow indicates the existence of a polynomial simulation.
The conversions corresponding to arrows marked (a) and (b) can be easily obtained by squaring 
the number of the states. (c) derives {}from~(b) and~(d).
The (trivial) dashed arrow {}from oblivious, sweeping, and few reversal automata
to \twdfas\ are not depicted.
}\label{f:machines}
\end{center}%
\end{figure}
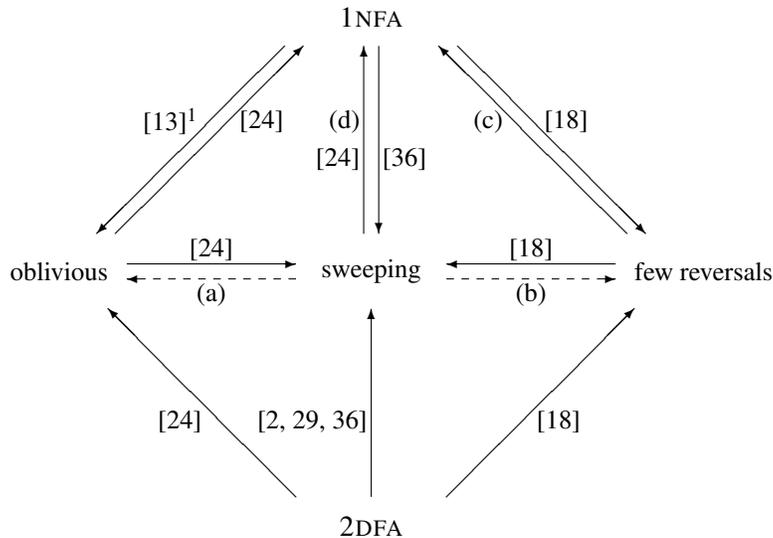

However, these results do not solve the general problem. In fact, it has been also proved that the simulations
of (unrestricted) \twdfas\ by these restricted models require exponentially many states.
See Figure~\ref{f:machines} for a summary of these and other separations.
Their proofs use rather involved arguments.

Concerning few reversals \twdfas, we already mentioned that a $o(n)$ upper bound on reversals implies a 
$O(1)$ upper bound~\cite{KP12b}. We can also compare the size of \twdfas\ making a fixed numbers of reversals.
For example, we observed that the language $I_n$ is accepted by a \twdfa\ with $n\plusc$ states that
makes only one reversal, while each \owdfa\ (i.e., each \twdfa\ making $0$ reversals) needs $2^n$ states
to accept it. Hence, \twdfas\ making $0$ reversals can be exponentially larger than
\twdfas\ making $1$ reversal.

\emph{What about \twdfas\ making $k$ versus \twdfas\ making $k+1$, for $k>0$?}

In the case $k=1$ this question has been solved by Balcerzak and Nivi\'nski~\cite{BN10}, by
proving an exponential separation. Recently Kapoutisis and Pighizzini extended this separation
to each integer $k$, providing an infinite reversal hierarchy of \twdfas~\cite{KP12b}.
It should be interesting to investigate similar questions in the nondeterministic case.

\footnotetext[\thetempcount]{A stronger separation can be given
by considering the degree of non-obliviousness, that counts
the number of  different trajectories of the head on inputs of the same length.
Hence, a \twdfa\ has a sublinear degree of non-obliviousness if and only if the number 
of different trajectories on inputs of length $n$ is $o(n)$.
In~\cite{HS03} it was proven that the simulation of \ownfas\ by \twdfas\ with a sublinear 
degree of non-obliviousness requires exponentially many states.}

\section{The Case of Unary Languages}

Unary languages are defined over a one letter alphabet $\Sigma$.
In the following we stipulate $\Sigma=\set{a}$.

\begin{figure}[hbt]
\begin{center}
\setlength{\unitlength}{0.5cm}
\begin{picture}(16,15)(10.5,9.5)
      \put(13,12){\makebox(0,0){\twdfa}}
      \put(13,22){\makebox(0,0){\owdfa}}
      \put(24,22){\makebox(0,0){\ownfa}}
      \put(24,12){\makebox(0,0){\twnfa}}

      \put(14.5,22.2){\vector(1,0){8}}
      \put(18.5,22.4){\makebox(0,0)[b]{$n$}}

      \put(15,21){\vector(1,-1){8}}
      \put(21.3,15){\makebox(0,0)[lb]{$n$}}

      \put(12.8,21){\vector(0,-1){8}}
      \put(12.6,17.5){\makebox(0,0)[rb]{$n$}}

      \put(22.5,21.8){\vector(-1,0){8}}
      \put(18.5,21.7){\makebox(0,0)[t]{$\Hn{n}$}}

      \put(25.5,21){\oval(2,2)[tr]}
      \put(26.5,21){\line(0,-1){9.5}}
      \put(24.5,11.5){\oval(4,4)[br]}
      \put(24.5,9.5){\line(-1,0){10.5}}
      \put(14,10.5){\oval(2,2)[bl]}
      \put(13,10.5){\vector(0,1){0.6}}
      \put(24,9.6){\makebox(0,0)[b]{$\Theta(n^2)$}}

      \put(24.2,21){\vector(0,-1){8}}
      \put(24.5,17.5){\makebox(0,0)[lb]{$n$}}

      \put(13.2,13){\vector(0,1){8}}
      \put(13.3,17.5){\makebox(0,0)[lb]{$\Hn{n}$}}

      \put(11.5,13){\oval(2,2)[bl]}
      \put(10.5,13){\line(0,1){9.5}}
      \put(12.5,22.5){\oval(4,4)[tl]}
      \put(12.5,24.5){\line(1,0){10.5}}
      \put(23,23.5){\oval(2,2)[tr]}
      \put(24,23.5){\vector(0,-1){0.6}}
      \put(14,24.4){\makebox(0,0)[t]{$\Hn{n}$}}

      \put(14.5,12.2){\vector(1,0){8}}
      \put(18.5,12.3){\makebox(0,0)[b]{$n$}}

      \put(22.5,13){\vector(-1,1){8}}
      \put(19.5,15){\makebox(0,0)[rb]{$\Hn{n}$}}

      \put(22.5,11.8){\vector(-1,0){8}}
      \put(18.5,11.5){\makebox(0,0)[t]{$?$}}

      \put(23.8,13){\vector(0,1){8}}
      \put(23.6,17.5){\makebox(0,0)[rb]{$\Hn{n}$}}

\end{picture}
\caption{Costs of the {\em optimal} simulations between different kinds of
{\em unary} automata. An arc labeled $f(n)$ 
{}from a vertex $x$ to a vertex $y$ means that a unary $n$-state automaton
in the class $x$ can be simulated by an $f(n)$-state automaton in the class $y$.
The $\Hn{n}$ costs for the simulations of \ownfas\ and \twdfas\ by
\owdfas\ as well the cost $\Theta(n^2)$ for the simulation of \ownfas\ by \twdfas\ have been
proved in~\protect{\cite{Chr86}}. The~$\Hn{n}$ cost for the simulation
of \twnfas\ and \owdfas\ has been proved in~\protect{\cite{MP01}}.
The other $\Hn{n}$ costs are easy consequences. All the $n$ costs are trivial.
The arc labeled ``?'' represents the open question of Sakoda and Sipser.}
\label{f:unary}
\end{center}
\end{figure}
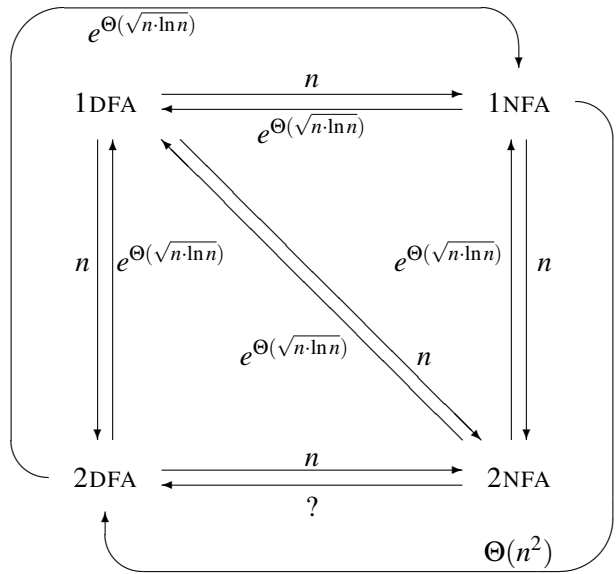

The state costs of the optimal simulations between different variant of
unary automata have been obtained by Chrobak~\cite{Chr86} and by Mereghetti and
Pighizzini~\cite{MP01} and are summarized in Figure~\ref{f:unary}.

{}From the picture we can observe that the cost of the optimal simulations in the unary
case can be smaller than in the general case. For example the cost of the simulation
of $n$-state \ownfas\ reduces {}from~$2^n$  to~$\Hn{n}$. Quite surprisingly, eliminating
at the same time both nondeterminism and two-way motion costs as eliminating only one of them.

The question \ownfas\ versus \twdfas\ has been solved in the unary case
in~\cite{Chr86} by showing that the tight cost is polynomial, more precisely $\Theta(n^2)$.
This gives also the best known lower bound for the general case.

In spite the  unary case looks simpler than the general one,
the question of \twnfas\ versus \twdfas\ not only is still open even in
this case, but it seems also to be difficult and, at the same time, very challenging.
We will now discuss its status.

\subsection*{Normal Forms for Unary Nondeterministic Automata}

The ``simplicity'' of automata over a unary alphabet, with respect to automata over a general
alphabet, allows to give normal forms for unary \ownfas\ and \twnfas. These forms, at the price of
a small increasing in the number of the states, strongly restrict the use of nondeterminism and
head reversals.

For the one-way case we mention the Chrobak normal form~\cite{Chr86}. In this form the transition
graph of the automaton consists of a deterministic path {}from the initial state to a state $q$, together
with $k\geq 0$
deterministic loops. {}From the state $q$ there are $k$ outgoing edges, each one of them connects
$q$ to exactly one state in each of the $k$ loops. Hence, a \ownfa\ in this form is allowed to make in its
computation at most one nondeterministic choice, when it is in the state $q$.
A degenerate case of \ownfa\ in Chrobak normal form is an automaton whose transition graph consists
exactly of one deterministic loop, without the initial path.
Each $n$-state unary \ownfa\ can be converted into an equivalent one in Chrobak normal form
with no more than $n^2$ states in the initial path and $n$ states in the loops. Hence the conversion does
not significantly increase the number of the states.\footnote{Besides~\cite{Chr86},
we refer the reader to~\cite{Gaw11, Gef07, Saw10} . All these papers present different
algorithms and techniques for the conversion of unary \ownfas\ into Chrobak normal form.}

A generalization of the Chrobak normal form to the two-way case has been obtained by Geffert, Mereghetti,
and Pighizzini~\cite{GMP03}.
In order to present it, it is useful to relax the notion of equivalence between automata, by
allowing a finite number of ``errors''. More precisely, two finite automata are said to be 
\emph{almost equivalent} if the symmetric difference of their accepted languages is finite,
i.e., the languages accepted by the two automata coincide expect for a finite number of strings.

\begin{theorem}[\cite{GMP03}]\label{t:nform}
  Each $n$-state unary \twnfa\ $A$ can be transformed into an almost equivalent \twnfa~$M$ such that
  \begin{itemize}
  \item $M$ is quasi-sweeping, namely, head reversals and nondeterministic choices are possible only when
  the head is scanning the endmarkers.\footnote{In~\cite{Sip80} the term \emph{sweeping}  was
  introduced for \emph{deterministic} automata making head reversals only at the endmarkers. It is natural
  to extend this notion to the nondeterministic case, to denote \twnfas\ making head reversals also at
  the endmarkers. In this case we have a further restriction: even nondeterministic decisions can taken
  only when the input head is scanning the endmarkers, not on ``real'' input symbols.}
  \item $M$ has at most $2n+2$ states,
  \item the languages accepted by $A$ and $M$ can differ only on strings of length at most $5n^2$.
  \end{itemize}
\end{theorem}
An inspection to the proof of Theorem~\ref{t:nform} shows that $M$ and its computations have a very
simple structure (see also~\cite{GP11}). In particular, in each traversal of the input $M$ uses a
deterministic loop to count the input length modulo one integer.

The \twnfa\ $M$ can be easily turned into an automaton ``fully'' equivalent to the original
\twnfa\ $A$, by adding $5n^2\plusc$ states, used to fix, in a preliminary scan of the input,
the ``errors''.

We point out that for unary \twdfas\ a similar normal form has been obtained in~\cite{KO11}.

The normal form in Theorem~\ref{t:nform} gives a strong simplification of 
unary \twnfas\ which has been an important tool to prove several
results on unary \twnfas.
First of all, it has been used in~\cite{GMP03} to prove a subexponential, but still superpolynomial
upper bound for the conversion of unary \twnfas\ into equivalent~\twdfas:

\begin{theorem}[\cite{GMP03}]\label{t:unaryconversion}
  Each unary $n$-state \twnfa\  can be simulated by a \twdfa\  with $e^{O(\ln^2n)}$ states.
\end{theorem}

It is interesting to discuss the main idea in the proof of this result.
Suppose the given $n$-state  \twnfa~$A$ is already in the normal form of Theorem~\ref{t:nform}.
We can observe that if an accepting computation $C$ visits the left endmarker more than $n$ times,
then there exists a shorter accepting computation $C'$ on the same input. In fact, in $C$ at least a same
state $q$ must be visited twice with the head at the left endmarker and so the computation $C'$ can be obtained by
cutting the part of $C$ between the two repetitions.
Hence, if we assume acceptance on the left endmarker, to detect if an input string is accepted it is 
enough to check the existence of a computation starting in 
the initial state with the head on the left endmarker,
ending in a final state with the head on the same endmarker, and visiting the left endmarker at most $n$ times.

To this aim we can introduce a predicate $reachable(p,q,k)$ which holds true exactly when there is a path 
starting in the state $p$ on the left endmarker, ending in the state $q$ on the same endmarker and
visiting it at most $k$ times.
This predicate can be recursively computed using a divide-and-conquere technique.
The implementation of the resulting procedure leads to a \twdfa\  with $e^{O(\ln^2n)}$ states.

In the case the given automaton is not in normal form, we first convert it into an
almost equivalent \twnfa\ in normal form and then we apply the above procedure to the resulting
automaton. Finally, with a small modification which does not increase the state upper bound,
we can fix the ``errors'', i.e., we can manage strings of length $\leq 5n^2$, in order to 
obtain a \twdfa\  fully equivalent to the original \twnfa.

\medskip

The upper bound in Theorem~\ref{t:unaryconversion} is subexponential, in the sense that it grows
less than the exponential function $e^n$, but it is superpolynomial, in fact it grows faster than any
polyomial.

The natural question is investigating whether or not it is tight.
At the moment we do not have an answer to it.
However, the question is related to the relationship
between deterministic and nondeterministic logarithmic space. 
The discussion of this point is postponed to the next section.

The normal form in Theorem~\ref{t:nform} has been used to prove other interesting properties
of unary \twnfas. Among them:

\begin{itemize}
	\item Each unary $n$-state \twnfa\ accepting a language $L$ can be transformed into a 
	\twnfa\ with $O(n^8)$ states accepting the complement of $L$~\cite{GMP07}.
	\item Each unary $n$-state \twnfa\ can be transformed into an equivalent
	\emph{unambiguous} \twnfa\ with a number of states polynomial in $n$~\cite{GP11}.
\end{itemize}
The proof of the first result is given by using an \emph{inductive counting} technique.
The second result was obtained adapting one of constructions discussed in the next section
(in particular, the construction used to prove Lemma~\ref{l:onlyif}).

\section{Relationships with the \logspace\ versus \nlogspace\ Question}

Interesting connections between the question of Sakoda and Sipser and the
open  question of the relationship between the classes of languages accepted
in logarithmic space by deterministic and nondeterministic Turing machines 
(denoted by \logspace\ and \nlogspace, respectively) have been obtained.
In this section we will briefly discuss them.

\begin{itemize}
\item[(i)] First of all, Berman and Lingas~\cite{BL77} proved that if $\logspace=\nlogspace$
then for each $n$-state \twnfa\ $A$ with an input alphabet of $\sigma$ symbols there exists
a \twnfa\ $B$ with a number of states polynomial in $n$ and $\sigma$ which agrees with $A$
on strings of length at most $n$. Hence $\logspace=\nlogspace$ implies a
polynomial simulation of \twnfas\ by \twdfas\ on ``short'' inputs.
\end{itemize}
This result was recently improved along the following lines.

\begin{itemize}
\item[(ii)] Geffert and Pighizzini~\cite{GP11} considered the unary case.
  They proved that $\logspace=\nlogspace$ would imply a polynomial simulation of unary \twnfas\
  by \twdfas.\footnote{The restriction to the unary case concerns only two-way
  automata, not the classes $\logspace$ and $\nlogspace$.} 
  Compared with condition (i), we can observe that while only devices with a unary input alphabet
  are considered here, the restriction on the length of the inputs is removed.

  This result shows the relevance of the unary case. In fact, proving
  the optimality of the bound in Theorem~\ref{t:unaryconversion} or even proving a smaller but
  still  superpolynomial lower bound for the simulation of unary \twnfas\ by \twdfas\ would
  imply the separation of $\logspace$ and $\nlogspace$.
    
\item[(iii)] Kapoutsis~\cite{Kap11a} generalized the condition (i) by 
  proving that $\lpoly\supseteq\nlogspace$ if and only if for each $n$-state \twnfa\ 
  $A$ with an input alphabet of $\sigma$ symbols there exists
  a \twnfa\ $B$ with a number of states polynomial in $n$ which agrees with $A$
  on strings of length at most $n$, where $\lpoly$ denotes the class of languages
  accepted by deterministic logspace bounded machines that can access a 
  \emph{polynomial advice}~\cite{KL82}.\footnote{A \emph{polynomial advice} is a sequence of strings 
  $(\alpha_n)_n\geq 0$, such that the length of $\alpha(n)$ is bounded by a polynomial in $n$.
  Together with an input string $x$, the machine receives the advice corresponding to
  the length of $x$, namely the string $\alpha(|x|)$.}
	Hence $\lpoly\supseteq\nlogspace$ is \emph{equivalent}
  to the existence of a state polynomial simulation of \twnfas\ by \twdfas\ on ``short'' inputs. 
  Since $\lpoly\supseteq\logspace$ and $\logspace\subseteq\nlogspace$, the \emph{only-if} 
  condition is stronger than the condition (i). 
  Furthermore, in this case the converse also holds.
  
\item[(iv)] Quite recently, Kapoutsis and Pighizzini~\cite{KP12a} proved the equivalence
  between $\lpoly\supseteq\nlogspace$ and several other propositions. In particular,
  they show that $\lpoly\supseteq\nlogspace$ is \emph{equivalent}
  to the existence of a state polynomial simulation of \emph{unary} \twnfas\ by \twdfas.
  As for (iii), we can observe that the \emph{only-if} condition is stronger
  than the condition in (ii) and, furthermore, in this case also the converse holds.
  
\end{itemize}
We are now go to discussing more into details (ii) and (iv).

\subsection*{The Graph Accessibility Problem}

A central role in the above mentioned investigations of the relationships between the 
$\logspace$ versus $\nlogspace$ and $\lpoly$ versus $\nlogspace$ questions 
and the problem of Sakoda and Sipser in the unary case
is played by the \emph{Graph Accessibility Problem} (\gap),
which is the problem of deciding given directed graph $G=(V,E)$ and
two fixed vertices $s,t\in V$, whether or not there exists a path {}from
$s$ to~$t$.\footnote{As customary, we use \gap\ also to denote the set of
positive instances of the graph accessibility problem. Hence, we write~$G\in\gap$ 
if and only if the given directed graph $G$ contains a path
connecting two (implicitly) fixed vertices $s$ and $t$.}

It is well known that \gap\ is an \nlogspace-complete problem~\cite{Sav70}. 
Hence, $\gap\in\nlogspace$ and, moreover, $\gap\in\logspace$ if and only if
$\logspace=\nlogspace$. In other words, this means that \gap\ is an hardest
problem in \nlogspace. As we discuss below, the restriction of \gap\ to a fixed
set of vertices represents in some sense (and under a suitable encoding) an hardest
language for unary \twnfas.

First of all, in~\cite{GP11} it was shown how to reduce the language
accepted by a unary $n$-state \twnfa\ $A$ to a graph with $N=O(n)$ vertices.
In other words, given an integer $m$ it is possible to obtain a graph
$G(m)$ with $N$ vertices such that the unary string $a^m$ is accepted by $A$ if and only if
$G(m)\in\gap$. Furthermore, the reduction can be computed by a finite state
transducer of size polynomial in $N$.

\begin{figure}[hbt]
\begin{center}
\setlength{\unitlength}{0.275cm}%
  \begin{picture}(23,6)(0,0.5)
  \put(0,3){\vector(1,0){4}}
  \put(1,4){\makebox(0,0){$a^m$}}
  \put(4,1){\framebox(4,4){$G$}}
  \put(11,4){\makebox(0,0){$G(m)$}}
  \put(8,3){\vector(1,0){6}}
  \put(14,1){\framebox(4,4){$\dgap$}}
  \put(18.05,3.2){\vector(2, 1){3}}
  \put(21.2,4.7){\makebox(0,0)[l]{yes}}
  \put(18.05,2.8){\vector(2,-1){3}}
  \put(21.2,1.3){\makebox(0,0)[l]{no}}
  \put(3.3,0.5){\framebox(15.2,5){ }}
  \end{picture}
  \caption{Simulating a unary \twnfa\ with a \twdfa\ of polynomial size,
  under the hypothesis $\logspace=\nlogspace$}
  \label{f:dgap}
\end{center}
\end{figure}
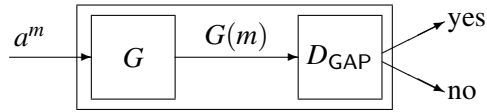

If $\logspace=\nlogspace$ then there is a logspace bounded deterministic machine
that solves \gap. By restricting this machine to inputs encoding graphs with $N$ 
vertices, we obtain a finite state automaton \dgap\ which can  decide whether
or not the graph $G(m)$ resulting {}from the above reduction is in \gap.
By a suitable composition of the transducer with \dgap\ we get a
\twdfa\ $B$ equivalent to the original \twnfa\ $A$, with a number of states polynomial 
in $n$, the number of states of $A$ (see Figure~\ref{f:dgap}).
We address the reader to~\cite{GP11} for details. In particular we point out that the
reduction uses the normal form for unary \twnfas\ presented in Theorem~\ref{t:nform}.
This construction has been extended to outer nondeterministic automata in~\cite{GGP12}.
Furthermore, with a similar technique, it is possible to show that unary
\twnfas\ and \twonfas\ over any input alphabet can be simulated by equivalent \emph{unambiguous}
\twnfas\ with polynomially many states~\cite{GGP12,GP11}.\footnote{These simulations
do not require the assumption $\logspace=\nlogspace$.}

It is quite natural to ask if the converse also holds, i.e., if a state
polynomial simulation of unary \twnfas\ by \twdfas\ would imply $\logspace=\nlogspace$.
The main problem in trying to prove such a result is related to the uniformity.
In particular, in~\cite{GP11} it is proved even a stronger result, however using the additional
hypothesis that the conversion {}from unary \twnfas\ to \twdfas\ is computed by
a logspace bounded transducer.

On the other hand, it is not difficult to observe that the above described construction
works even under the weaker hypothesis $\lpoly\supseteq\nlogspace$,
i.e.:
\begin{lemma}
\label{l:onlyif}
  If\ $\lpoly\!\supseteq\!\nlogspace$ then the state cost of the simulation of unary
  \twnfas\ by \twdfas\ is polynomial.
\end{lemma}

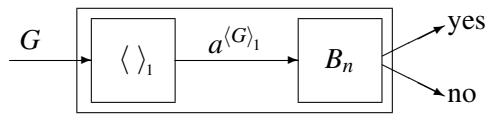
\begin{figure}[hbt]
\begin{center}
  \setlength{\unitlength}{0.275cm}
  \begin{picture}(23,5)(0,0.5)
  
  \put(0,3){\vector(1,0){4}}
  \put(1,4){\makebox(0,0){$G$}}
  
  \put(4,1){\framebox(4,4){$_{~}\uencoding{~}$}}
  
  \put(11,4){\makebox(0,0){$a^{\uencoding{G}}$}}
  \put(8,3){\vector(1,0){6}}
    
  \put(14,1){\framebox(4,4){$B_n$}}
  
  \put(18.05,3.2){\vector(2, 1){3}}
  \put(21.2,4.7){\makebox(0,0)[l]{yes}}
  \put(18.05,2.8){\vector(2,-1){3}}
  \put(21.2,1.3){\makebox(0,0)[l]{no}}
  
  \put(3.3,0.5){\framebox(15.2,5){ }}
 
  \end{picture}
  \caption{The machine $M$ solving \gap\ using $B_n$ as advice}
  \label{f:dfaGap}
\end{center}
\end{figure}

In~\cite{KP12a}, also the converse of Lemma~\ref{l:onlyif} has been proved.
The main idea is to exhibit, under the hypothesis that the state cost of the 
simulation of unary \twnfas\ by \twdfas\ is polynomial, a logspace bounded deterministic
machine $M$ which, making use of a polynomial advice, solves the graph accessibility problem.
This is done by the following steps:
\begin{itemize}
\item A function $\uencoding{~}$ mapping instances of \gap\ to unary strings is provided.
  For each integer $n$, the function $\uencoding{~}$ is a reduction {}from \gap\
  restricted to graphs with $n$ vertices to a unary language \Unarygap{n}.
\item A unary \twnfa\ $A_n$ recognizing \Unarygap{n} with a number of states polynomial in $n$
  is described.
\item The automaton $A_n$ is replaced by an equivalent \twdfa\ $B_n$.
\item An instance of \gap\ can be solved by combining the machine computing the reduction with 
  the \twdfa\ $B_n$, where $n$ is the number of vertices in the instance under consideration
  (hence $n$ depends only on the input length), see Figure~\ref{f:dfaGap}.
  In particular, the resulting machine $M$ receives the input string, which represents a graph~$G$, together
  with an encoding of the appropriate \twdfa\ $B_n$, where $n$ is the number of vertices of $G$.
  If the state cost of the simulation of unary \twnfas\ by \twdfas\ is polynomial then $B_n$
  can be encoded by a string of polynomial length in $n$. Such encoding is \emph{the polynomial advice for 
  $M$.} Furthermore, using a suitable encoding for \Unarygap{n} (we sketch some ideas below)
  the workspace used by $M$ can be bounded by a logarithmic function in~$n$.
\end{itemize}

We are going to describe the encoding $\uencoding{~}$ and the languages $\Unarygap{n}$.

For each integer $n$, let $V_n=\set{0,1,\ldots,n-1}$ and $K_n$ be the complete graph with 
vertex set $V_n$. With each edge $(i,j)$ of $V_n$ we associate a different prime $p_{(i,j)}$.
To this aim we choose the first $n^2$ prime numbers.

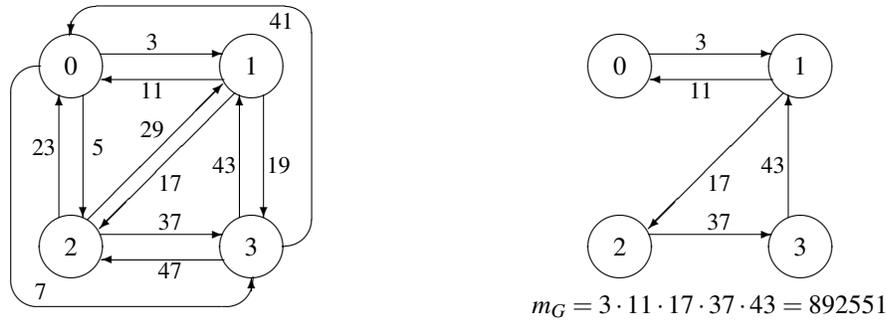
\begin{figure}[bt]
\begin{center}

\setlength{\unitlength}{0.4cm}
\begin{picture}(28,11)(0,-1)
\put(0,0){%
\begin{picture}(10,10)
	\put(2,2){\circle{2}}
      \put(2,2){\makebox(0,0){\small $2$}}	
	\put(8,2){\circle{2}}
      \put(8,2){\makebox(0,0){\small $3$}}	
	\put(2,8){\circle{2}}
      \put(2,8){\makebox(0,0){\small $0$}}	
	\put(8,8){\circle{2}}
      \put(8,8){\makebox(0,0){\small $1$}}	
	\put(2.95,2.4){\vector(1,0){4.1}}
	\put(7.1,1.55){\vector(-1,0){4.1}}
	\put(2.95,8.4){\vector(1,0){4.1}}
	\put(7.1,7.55){\vector(-1,0){4.1}}
	\put(1.6,2.95){\vector(0,1){4.1}}
	\put(2.4,7.05){\vector(0,-1){4.1}}
	\put(7.6,2.95){\vector(0,1){4.1}}
	\put(8.4,7.05){\vector(0,-1){4.1}}
	\put(2.55,2.9){\vector(1,1){4.5}}
	\put(7.45,7.1){\vector(-1,-1){4.5}}
	\put(1,4){\oval(2,8)[l]}
	\put(1,0){\line(1,0){6}}
	\put(7,1){\oval(2.1,2)[br]}
	\put(8,0.5){\vector(0,1){0.45}}
	\put(9,6){\oval(2,8)[r]}
	\put(3,10){\line(1,0){6}}
	\put(3,9){\oval(2.1,2)[tl]}
	\put(2,9.5){\vector(0,-1){0.45}}
	
	\put(4.7,7.2){\makebox(0,0){\footnotesize $11$}}
	\put(4.7,8.8){\makebox(0,0){\footnotesize $3$}}
	\put(5.3,1.2){\makebox(0,0){\footnotesize $47$}}
	\put(5.3,2.8){\makebox(0,0){\footnotesize $37$}}
	\put(7.1,4.7){\makebox(0,0){\footnotesize $43$}}
	\put(8.9,4.7){\makebox(0,0){\footnotesize $19$}}
	\put(1.1,5.3){\makebox(0,0){\footnotesize $23$}}
	\put(2.9,5.3){\makebox(0,0){\footnotesize $5$}}
	\put(4.7,5.9){\makebox(0,0){\footnotesize $29$}}
	\put(5.3,4.1){\makebox(0,0){\footnotesize $17$}}
	\put(1,0.5){\makebox(0,0){\footnotesize $7$}}
	\put(9,9.5){\makebox(0,0){\footnotesize $41$}}

\end{picture}
}
\put(18,0){
\begin{picture}(10,10)
	\put(2,2){\circle{2}}
      \put(2,2){\makebox(0,0){\small $2$}}	
	\put(8,2){\circle{2}}
      \put(8,2){\makebox(0,0){\small $3$}}	
	\put(2,8){\circle{2}}
      \put(2,8){\makebox(0,0){\small $0$}}	
	\put(8,8){\circle{2}}
      \put(8,8){\makebox(0,0){\small $1$}}	
	\put(2.95,2.4){\vector(1,0){4.1}}
	\put(2.95,8.4){\vector(1,0){4.1}}
	\put(7.1,7.55){\vector(-1,0){4.1}}
	\put(7.6,2.95){\vector(0,1){4.1}}
	\put(7.45,7.1){\vector(-1,-1){4.5}}
	
	\put(4.7,7.2){\makebox(0,0){\footnotesize $11$}}
	\put(4.7,8.8){\makebox(0,0){\footnotesize $3$}}
	\put(5.3,2.8){\makebox(0,0){\footnotesize $37$}}
	\put(7.1,4.7){\makebox(0,0){\footnotesize $43$}}
	\put(5.3,4.1){\makebox(0,0){\footnotesize $17$}}
	
	\put(5,0){\makebox(0,0){\small $m_G=3\cdot 11\cdot 17\cdot 37\cdot 43=892551$}}
\end{picture}
}
\end{picture}
\caption{The complete graph $K_4$ with a subgraph $G$. The number $p_{(i,j)}$ 
  associated with the edge~$(i,j)$ is the $(i\cdot n+j+1)th$ prime number. 
  In $K_4$ the edges $(i,i)$ are not depicted.}
\label{f:graph}
\end{center}
\end{figure}

A graph $G=(V_n,E)$ with $n$ vertices is encoded as the product 
of all prime powers corresponding to the edges in $E$ (see Figure~\ref{f:graph}), i.e., by the number 
\[\uencoding{G}=\prod_{(i,j)\in E}p_{(i,j)}\]
Conversely, with each integer $m$ we associate the graph $K_n(m)=(V_n,E(m))$
such that $(i,j)\in E(m)$ if and only if $p_{(i,j)}$ divides $m$.
It should be clear that $K_n(\uencoding{G}\!)=G$.

We can now define the \emph{unary encoding} of \gap\ restricted to graphs with $n$ vertices,
as the following language:
\[
\Unarygap{n}=\set{a^m\mid K_n(m)~\mbox{has a path {}from $0$ to $n-1$}}
\]

We now describe a \twnfa\ $A_n$ recognizing $\Unarygap{n}$.
Roughly speaking, $A_n$ implements the standard nondeterministic algorithm solving
\gap.
{}From a vertex $i$ (starting {}from $i\leftarrow 0$ at the beginning of the computation), 
$A_n$ guesses another vertex $j$ and then it verifies
whether $(i,j)\in E$. If this is the case, then $A_n$ continues the same simulation after making the
assignment $i\leftarrow j$,
up to reach $i = n-1$. However, if in a step a pair $(i,j)\notin E$ is reached, then $A_n$
hangs and rejects.
To check the condition $(i,j)\in E$, $A_n$ computes the length of its input modulo $p_{(i,j)}$.

\noindent
More into details:
\begin{itemize}
\item $A_n$ is outer nondeterministic and sweeping, i.e., it can reverse the input head direction
and make nondeterministic choices \emph{only} when the head is scanning one of the endmarkers.
Furthermore, in each traversal $A_n$ counts the input length modulo a prime number.
\item On the endmarkers each state is interpreted either as a copy of a vertex 
in $V_n$ or as an \emph{hang} state.
\item The automaton can traverse an input $a^m$ {}from one endmarker in a copy of vertex $i$
to the opposite endmarker in some copy of vertex $j$, without visiting the endmarkers in
between, if and only if the number $p_{(i,j)}$ divides $m$.
In particular, when the automaton is visiting one endmarker in a state representing the 
vertex $i$, it guesses another vertex $j$, by  entering an appropriate loop where it traverses and counts
the input modulo $p_{(i,j)}$. The state in this loop which corresponds to the remainder $0$ 
is interpreted as the vertex $j$ of the graph, the other states are interpreted as \emph{hang} states.
Hence, when the input head reaches the opposite endmarker, the automaton continues the simulation
or hangs and rejects depending on the reached state.
\item The computation starts on the left endmarker in a state representing the vertex $0$.
\item When a state representing the vertex $n-1$ is reached with the head on one of the endmarkers,
the automaton $A_n$ moves to an accepting state and stops the computation.
\end{itemize}
Using the properties related to the distribution of prime numbers, it can be proved that
\emph{the number of states of $A_n$ is polynomial in $n$.}

\medskip

Finally, we have to show that the machine $M$ works in logarithmic space.
Actually, we can observe that this is not true if we directly implement $M$ as in 
Figure~\ref{f:dfaGap}.
In fact the length of the unary encoding of a graph with $n$ vertices can be exponential
in $n$. For instance,
$\uencoding{K_n}$, the unary encoding of the complete graph of $n$ vertices, is the
product of first $n$ prime numbers, which is exponential in $n$.

This problem is solved as follows:

\begin{itemize}
\item The unary encoding is replaced by a ``prime encoding'' that, in this
  case, is a list of all primes associated with the edges in the input graph.
  Hence, the output of the reduction is this list.\footnote{More in general,
  a \emph{prime encoding} of a unary string $a^m$ is a sequence of the
  form $z_1\#z_2\#\cdots\#z_k$ where $z_1, z_2,\ldots, z_k$ are
  strings encoding in an \emph{arbitrary order} the prime powers in
  the factorization of $m$.}
\item Due to a structural property of \twdfas\ (see~\cite{KO11}), it is possible
  to modify the automaton $B_n$, still keeping polynomial its number of states,
  by replacing its unary input tape, with a tape containing a prime encoding of
  the unary input. Hence, after these modifications, the machine $M$ still
  solves \gap.
\item To be stored, the prime encoding would require polynomial space, which
  is still too much for our purposes. To avoid this problem, the prime encoding is
  not kept in the internal memory of $M$, but it is computed and recomputed
  ``on fly'', each time $B_n$ needs to access it. This is done 
  by restarting the machine that {}from the input graph $G$ computes the prime encoding.
\end{itemize}

Along these lines the converse of Lemma~\ref{l:onlyif} is proved.
This allows to obtain the following:

\begin{theorem}
\label{t:onlyif}
  $\lpoly\supseteq\nlogspace$ if and only if the state cost of the simulation of unary
  \twnfas\ by \twdfas\ is polynomial.
\end{theorem}
We address the reader to~\cite{KP12a} for the details and for the equivalence of $\lpoly\supseteq\nlogspace$ with several other statements.

\section{Concluding Remarks}

We strongly believe that the Sakoda and Sipser question is a very challenging problem which
deserves further investigation.
Several interesting models have been considered and many deep results have been obtained 
in the researches related to this question.
As pointed out, connections with space complexity have been discovered. This is not limited to
the relationships with the question of the power of nondeterminism in logspace bounded computations. In fact,
in more than one case, techniques {}from space complexity turn to be useful to study two-way
automata. For instance, the divide-and-conquere technique used to prove Theorem~\ref{t:unaryconversion}
derives {}from the proof of the famous Savitch Theorem~\cite{Sav70}.
The inductive counting tecnique used in~\cite{GMP07} to obtain the polynomial
complementation of \twnfas\ derives {}from the argument used to prove the closure
under complementation of nondeterministic space, the famous result independently
proved in 1988 by Immerman~\cite{Imm88} and Szelepcs{\'e}nyi~\cite{Sze88}.

Actually, the complexity theory for finite automata can be developed as a part of standard
complexity theory for Turing machines, with classes, reductions, complete problems and so
on. This approach was suggested in the original paper by Sakoda and Sipser~\cite{SS78}.
We recommend  the recent paper by Kapoutisis~\cite{Kap12} to the interested reader, where 
the name \emph{minicomplexity} is suggested for this theory.
The same author is working to collect and organize in a website 
all the material and the results in this area, see~\verb+www.minicomplexity.org+.

\bibliographystyle{eptcs}
\bibliography{biblioGio}
\end{document}